\documentclass[conference]{IEEEtran}

\pdfoutput=1

\IEEEoverridecommandlockouts

\usepackage[utf8]{inputenc}
\usepackage[english]{babel}

\usepackage[nocompress]{cite}
\usepackage{graphicx}
\usepackage[para]{threeparttable}
\usepackage{multirow}
\usepackage{multicol}
\usepackage{xcolor}
\usepackage{colortbl}
\usepackage[binary-units]{siunitx}
\usepackage{booktabs}
\usepackage{tabularx}
\usepackage[acronym]{glossaries}
\usepackage[draft=false, hidelinks]{hyperref}
\usepackage[printwatermark]{xwatermark}
\usepackage[capitalise]{cleveref}
\usepackage{subcaption}
\usepackage{etoolbox}
\usepackage{mathdots}
\usepackage{enumitem}
\usepackage{tikz}
\usepackage{adjustbox}
\usepackage[a4paper, total={184mm,239mm}]{geometry}
\usepackage{float}

\graphicspath{{figures/}}

\newcommand*\circled[1]{\unskip~\tikz[inner sep=.15ex,baseline=-.75ex)]{
    \node[circle,draw,fill=black,text=white] (char) {\textbf{\small#1}};}}

\newacronym{hpc}{HPC}{high performance computing}
\newacronym{isa}{ISA}{instruction set architecture}
\newacronym{fpu}{FPU}{floating-point unit}
\newacronym{fp}{FP}{floating-point}
\newacronym{mac}{MAC}{multiply-accumulate}
\newacronym{muldiv}{MULDIV}{integer multiply-divide unit}
\newacronym{alu}{ALU}{arithmetic logic unit}
\newacronym{fu}{FU}{functional unit}

\newacronym[firstplural=Core Complexes (CCs)]{cc}{CC}{core complex}
\newacronym{dmcc}{DMCC}{data movement CC}
\newacronym{fpss}{FPSS}{floating-point subsystem}
\newacronym{ssr}{SSR}{stream semantic register}
\newacronym{issr}{ISSR}{indirection stream semantic register}
\newacronym{tcdm}{TCDM}{tightly-coupled data memory}
\newacronym{frep}{FREP}{floating-point repetition}
\newacronym{dma}{DMA}{direct memory access}

\newacronym{csf}{CSF}{compressed sparse fiber}
\newacronym{csr}{CSR}{compressed sparse rows}
\newacronym{csc}{CSC}{compressed sparse columns}
\newacronym{dcsr}{DCSR}{doubly-compressed sparse rows}
\newacronym{sdcsr}{sDCSR}{sliced doubly-compressed sparse rows}
\newacronym{spmv}{SpMV}{sparse matrix-vector multiplication}
\newacronym{sdmv}{SDMV}{sDCSR matrix-vector multiplication}

\newacronym{spvv}{SpVV}{sparse vector-vector multiplication}
\newacronym{csrmv}{CsrMV}{CSR matrix-vector multiplication}
\newacronym{csrmm}{CsrMM}{CSR matrix-matrix multiplication}
\newacronym{cvr}{CVR}{compressed vectorization-oriented sparse row}
\newacronym{dl}{DL}{deep learning}

\newacronym{ml}{ML}{machine learning}
\newacronym{simd}{SIMD}{single-instruction, multiple-data}
\newacronym{sm}{SM}{streaming multiprocessor}
\newacronym{knl}{KNL}{Knights Landing}

\DeclareSIUnit\GE{GE}
\DeclareSIUnit\kGE{\kilo\GE}
\DeclareSIUnit\MGE{\mega\GE}
\DeclareSIUnit\flop{flop}
\DeclareSIUnit\flops{\flop\per\second}
\DeclareSIUnit\Gflops{\giga\flops}
\DeclareSIUnit\Tflops{\tera\flops}

\newfloat{listing}{htbp}{lst}
\floatname{listing}{Listing}

\setlist[description]{font=\normalfont\emph,labelindent=\parindent}

\newcommand{\x}{$\times$}

\IEEEaftertitletext{\vspace{-0.3\baselineskip}}

\begin{document}

\title{Indirection Stream Semantic Register Architecture\\ for Efficient Sparse-Dense Linear Algebra}

\author{
    \IEEEauthorblockN{%
    Paul Scheffler\textsuperscript{\textasteriskcentered}, %
    Florian Zaruba\textsuperscript{\textasteriskcentered}, %
    Fabian Schuiki\textsuperscript{\textasteriskcentered}, %
    Torsten Hoefler\textsuperscript{\textdagger}, %
    Luca Benini\textsuperscript{\textasteriskcentered}%
    }
    \IEEEauthorblockA{
        \textasteriskcentered~\textit{Integrated Systems Laboratory, ETH Zurich}, Switzerland \\
        \textdagger~\textit{Scalable Parallel Computing Laboratory, ETH Zurich}, Switzerland \\
        \{paulsc,zarubaf,fschuiki\}@iis.ee.ethz.ch, htor@inf.ethz.ch, lbenini@iis.ee.ethz.ch
    }
}

%\author{\textit{Authors omitted for blind review}}

\maketitle

\begin{abstract}
    Sparse-dense linear algebra is crucial in many domains, but challenging to handle efficiently on CPUs, GPUs, and accelerators alike; multiplications with sparse formats like CSR and CSF require indirect memory lookups. In this work, we enhance a memory-streaming RISC-V ISA extension to accelerate sparse-dense products through streaming indirection. We present efficient dot, matrix-vector, and matrix-matrix product kernels using our hardware, enabling single-core FPU utilizations of up to 80\% and speedups of up to 7.2x over an optimized baseline without extensions. A matrix-vector implementation on a multi-core cluster is up to 5.8x faster and 2.7x more energy-efficient with our kernels than an optimized baseline. We propose further uses for our indirection hardware, such as scatter-gather operations and codebook decoding, and compare our work to state-of-the-art CPU, GPU, and accelerator approaches, measuring a 2.8x higher peak FP64 utilization in CSR matrix-vector multiplication than a GTX 1080 Ti GPU running a cuSPARSE kernel.
\end{abstract}


%%%%%%%%%%%%%%%%%
%%   CONTENT   %%
%%%%%%%%%%%%%%%%%

\section{Introduction}
\label{sec:intro}

Sparse vector, matrix, and tensor operations pose a formidable challenge for today's processor architectures, which are optimized towards highly regular and vectorizable workloads.
In sparse tensors, a significant fraction of elements is zero.
A popular approach to reducing the runtime and memory requirements of sparse multiplications is to skip these zeros and process only nonzero elements.
Consider the \gls{csr} format, which represents the rows of a sparse matrix \verb|A| as a list of \emph{nonzeros} \verb|A_vals| and a list of their positions \verb|A_idcs|.
In this format, \gls{spmv} can be implemented as follows:
\begin{adjustbox}{nofloat=figure,vspace=\medskipamount}
\hspace{\parindent}%
\includegraphics{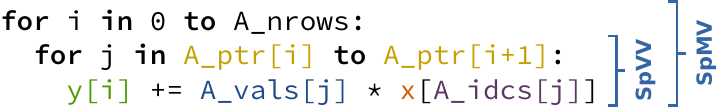}\hfill
\end{adjustbox}\noindent
where \verb|A_ptr| delimits the rows of \verb|A|. The \gls{spmv} consists of multiple \glspl{spvv}, one for each element of the result vector \verb|y|.
Let us therefore focus on \gls{spvv}, which compiles to the following RISC-V assembly:
\begin{adjustbox}{nofloat=figure,vspace=\medskipamount}
\hspace{\parindent}%
\includegraphics{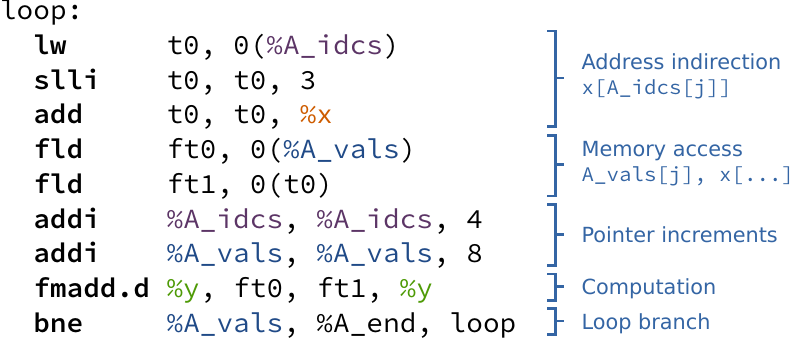}\hfill
\end{adjustbox}\noindent
On a single-issue core, each loop iteration takes at least nine cycles to execute. Only one instruction performs necessary computation, using the \gls{fpu} at most \SI{11}{\percent} of the time.
A superscalar core, even when ignoring dependencies, would need nine issue slots to achieve full \gls{fpu} utilization.

The main issue is that the \emph{indirection} \verb|x[A_idcs[j]]|, which accesses elements of the dense vector \verb|x|, requires multiple integer instructions.
The low \gls{fpu} utilization a simple core achieves, and the significant instruction frontend complexity a superscalar core requires, severely limit our ability to execute sparse-dense operations in an area- and energy-efficient manner.

Efficient processing of sparse data structures is essential as sparsity is ubiquitous in many domains~\cite{zhang2015survey}; the SuiteSparse Matrix collection~\cite{davis2011suitesparse} curates sparse matrices from various real-world problems, covering physical simulation, mathematics, computer science, biology, and economics among other fields.
Moreover, sparsification techniques in \gls{ml} can significantly reduce the computational footprint incurred to attain a given accuracy \cite{gale2019state}.
These frontiers motivate new computer architectures that handle sparsity efficiently.

Existing superscalar out-of-order architectures struggle significantly with the high control-to-compute ratio inherent to sparse formats, reaching only a fraction of their peak compute throughput.
Consider the Intel Xeon Phi 7250~\cite{xie2018cvr}: with a highly optimized machine-specific matrix compression scheme, it achieves \SI{21}{\Gflops}, a mere \SI{0.7}{\percent} of its peak compute.
Even worse, the extensive out-of-order capabilities leveraged are not free: at a thermal design power of \SI{230}{\watt} and die sizes exceeding \SI{500}{\milli\meter\squared}, such a processor exhibits very low area and energy efficiencies.
To achieve high efficiency, it is crucial to maximize the silicon area and energy dedicated to \glspl{fpu}, which do the useful work, and that they are highly utilized.

\glsunset{issr}
In this paper, we propose \emph{indirection stream semantic registers} (ISSRs), an architectural extension to in-order, single-issue cores enabling high \gls{fpu} utilization in sparse-dense multiplication \emph{without} major impact on microarchitectural complexity.
\Glspl{issr} are based on Schuiki et al.'s \glspl{ssr}~\cite{schuiki2019ssr}, which offer a way to reach close to 100\% \gls{fpu} utilization with single-issue cores on dense data structures.
This is achieved by allowing register reads/writes to implicitly encode a load/store operation together with an associated address computation.
Zaruba et al.'s Snitch~\cite{zaruba2020snitch} shows that an extremely small \SI{10}{\kGE} core can fully utilize a large double-precision \SI{100}{\kGE} \gls{fpu} when \glspl{ssr} are combined with \gls{frep} hardware loops, greatly boosting efficiency.
Our work's key rationale is extending \glspl{ssr} to support the indirection necessary for sparse-dense linear algebra, incurring only a small \SI{4.4}{\kGE} increase in hardware complexity.
This allows us to rewrite our \gls{spvv} example as
\begin{adjustbox}{nofloat=figure,vspace=\medskipamount}
\hspace{\parindent}%
\includegraphics{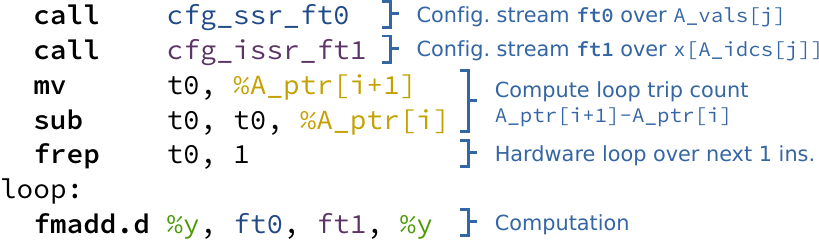}\hfill
\end{adjustbox}\noindent
where \verb|ft0| and \verb|ft1| now have stream semantics.
Since address calculation, indirection, and load/stores are now fully handled in hardware, the loop body consists of only the \verb|fmadd.d| instruction, which we execute using the \verb|frep| hardware loop.
Intuitively, the expected speedup is 9\x.
Since the indirection incurs additional memory bandwidth, the speedup achieved in practice depends on whether this traffic is multiplexed onto existing memory channels or whether additional channels are provided.
We will show that in the former case, a single core can reach up to \SI{80}{\percent} \gls{fpu} utilization on \gls{spmv} using real-world matrices, yielding a 7.2\x~peak speedup over hand-optimized RISC-V assembly and 5.6\x~over the best implementation using dense \glspl{ssr}.
In an eight-core cluster with added memory contention and parallelization overheads, our extension still achieves a 5.8\x~speedup over our RISC-V baseline.
To summarize, our contributions are as follows:

\begin{enumerate}
    \item A lightweight extension to \glspl{ssr} to handle streaming indirection, enabling highly efficient sparse-dense products in single-issue cores (\Cref{sec:arch}).
    \item A programming model for the proposed \glspl{issr} and corresponding efficient sparse-dense dot (SpVV), CSR matrix-vector (CsrMV), and CSR matrix-matrix (CsrMM) product implementations (\Cref{sec:prog}).
    \item Significant performance and energy efficiency benefits, achieving up to 7.2\x~speedups on a single core and 5.8\x~and 2.7\x~gains on an eight-core cluster, respectively. We evaluate the impact of \gls{issr} on area and power in a 22\,nm technology, increasing the area of an eight-core cluster with \glspl{ssr} by only \SI{0.8}{\percent} (\Cref{sec:results}).
    \item A comparison to state-of-the-art CPUs, GPUs, and sparse accelerator approaches, with \gls{issr} achieving 2.8\x~higher peak floating-point utilization than GPUs (\Cref{sec:relwork}).
\end{enumerate}

\section{Architecture}
\label{sec:arch}

Our indirection hardware extension is mostly confined to the \gls{ssr}'s \emph{address generator}; we will focus on its architecture before presenting the \gls{issr} streamer and its integration into the Snitch cluster~\cite{zaruba2020snitch} in which we evaluate our work. We note that \glspl{issr} are functionally orthogonal to this platform; they could be used with any core, instruction set, or hardware loop.

\subsection{Address Generator}
\label{sec:arch_agen}

\begin{figure}[t]
    \centering
    \includegraphics[width=0.845\linewidth]{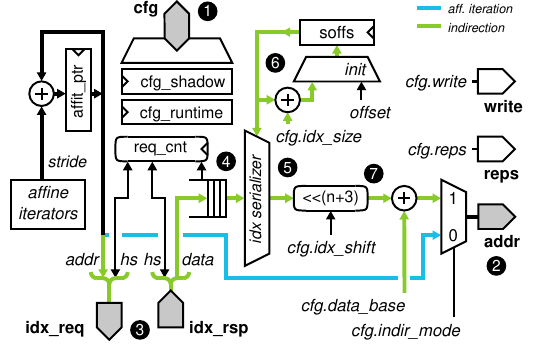}
    \caption{ISSR address generator architecture}
    \label{fig:arch_addr_gen}
\end{figure}

\cref{fig:arch_addr_gen} shows our extended \gls{issr} address generator. As in the \gls{ssr}, it exposes a \emph{shadowed} configuration register interface to the host core \circled{1}, allowing the setup of a new job while another is running, and a control interface to the data mover \circled{2}. It adds a read-only memory interface for index streaming \circled{3}.

The four nested \gls{ssr} affine address iterators are left unchanged: at each emitted datum, the stride of the outermost iterating loop is added onto a shared memory pointer.
In the existing \emph{affine iteration} mode, this pointer directly provides addresses to the \emph{streamed data} to the \gls{ssr}'s data mover. In the new \emph{indirection} mode, it provides addresses to the \emph{index array} indicating the data to be streamed; we must first fetch these indices from memory and add their corresponding offsets to a base address before emission to the data mover.

In indirection mode, we fix the affine iterator configuration to one dimension with a stride of eight bytes, loading a contiguous stream of 64-bit words from the shared, cluster-local \gls{tcdm} into a decoupling FIFO \circled{4}. An \emph{outstanding request counter} regulates address emission to prevent downstream blocking or a FIFO overflow.

Our hardware can read arrays of either 32-bit or 16-bit indices. To this end, an \emph{index serializer} \circled{5}, backed by a two-bit \emph{short offset} counter \circled{6}, extracts 16- or 32-bit indices from the buffered 64-bit index words. To simplify the programming model, arbitrary index array alignment is supported.

The indices are statically shifted to 64-bit word offsets to serve the double-precision FPU and added to the configured data base address \circled{7}. Optionally, they can be shifted further by a programmable offset, enabling indirection into higher-level axes of power-of-two-strided tensors or tiles. This offers an attractive tradeoff between arbitrary striding, requiring a hardware multiplier, and no higher-level iteration support.

As in \glspl{ssr}, the indices and addresses used in indirection have design-time-parameterizable widths between 16 and 32 bit, both defaulting to 18 bit to cover our \SI{256}{\kibi\byte} \gls{tcdm}.

\subsection{ISSR Streamer}
\label{sec:arch_streamer}

\begin{figure}[t]
    \centering
    \includegraphics[width=0.78\linewidth]{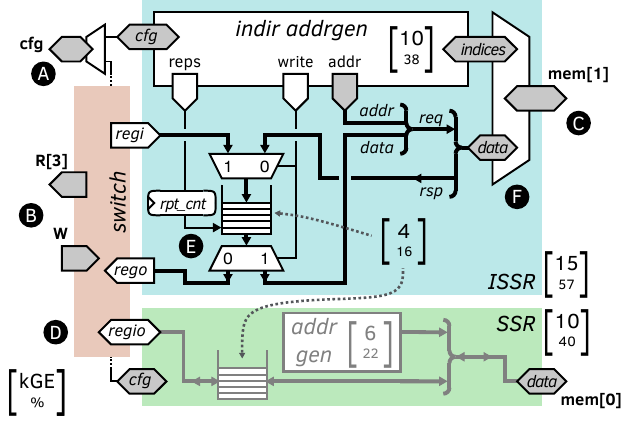}
    \caption{ISSR and ISSR streamer architecture}
    \label{fig:arch_hssr_streamer}
\end{figure}

\cref{fig:arch_hssr_streamer} shows the architecture of \gls{issr} and its containing \emph{streamer}. The streamer maintains the existing IO, exposing a shared configuration interface to the core \circled{A}, a register file interface to the FPU \circled{B}, and one port per \gls{ssr} to the memory system \circled{C}. The switch maps each \gls{ssr} to a specific architectural register while enabled \circled{D}.

Like the \gls{ssr}, the \gls{issr} provides a FIFO decoupling the register and memory streams, reused in both reading and writing \circled{E}. In our design, the \gls{issr}'s index and data memory ports are combined by a round-robin multiplexer~\circled{F}: for each 16- or 32-bit index word, two or four 64-bit words are fetched, yielding a peak data mover utilization of $2/3$ or $4/5$, respectively. This arbitration could also be done in the streamer or core, or omitted entirely by providing three ports per core, trading higher utilization and performance for approximately $1.5\times$ larger interconnect logic. In this work, we show an area-optimized extension with one port per \gls{ssr}, enabling direct comparisons to the existing two-port, two-\gls{ssr} streamer.

The presented streamer provides one \gls{issr} and one \gls{ssr}, but it could combine any number of either given sufficient memory ports. The new indirection semantic does not change exception handling, which is described in detail by Schuiki et al.\cite{schuiki2019ssr}.

\subsection{Core and Cluster Integration}
\label{sec:arch_cc_clust}

\begin{figure}[t!]
    \centering
    \includegraphics[width=1.0\linewidth]{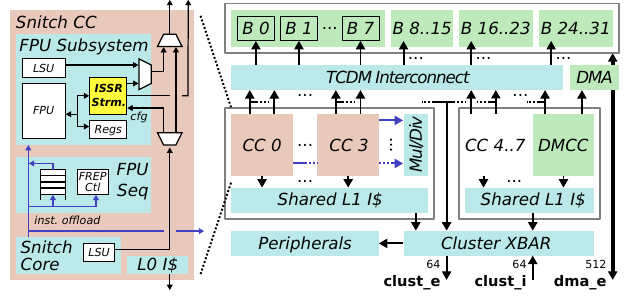}
    \caption{Snitch core complex (CC) and cluster architecture. Our ISSR streamer is highlighted in yellow and with bold font.}
    \label{fig:arch_cc_clust}
\end{figure}

\cref{fig:arch_cc_clust} illustrates the streamer's integration into the existing Snitch \gls{cc} and cluster. Like the \gls{ssr} streamer, it is part of the double-precision FPU subsystem in each \gls{cc} and multiplexed with the floating-point register file.

We kept the existing \gls{cc} memory topology, providing an exclusive port to the \gls{issr} while combining the core, FPU, and \gls{ssr} requests into another: this maximizes sparse-dense performance by giving the core opportunities for memory requests while the \gls{issr} is fetching indices without blocking the \gls{ssr} and impacting performance. Furthermore, this allows existing code to run without performance degradation.

The cluster contains eight worker \glspl{cc} organized into two \emph{hives}, sharing an L1 instruction cache and an integer multiply-divide unit. Our \gls{tcdm} has 32 banks totaling 256 \si{\kibi\byte}. A 512-bit \gls{dma} engine efficiently moves data blocks between the \gls{tcdm} and main memory~\cite{kurth2020manycore}. It is controlled by a lightweight \gls{dmcc} without an FPU, which can also be used for cluster coordination.
\section{Programming and Applications}
\label{sec:prog}

The \gls{issr}'s address stream is configured by the core through its memory-mapped register interface, which enables few-to-single-cycle setups. While future works could extend compilers to map indirections in high-level code to \glspl{issr}, we focus here on accelerating sparse-dense products on an instruction level.

\subsection{Accelerable Sparse Formats}
\label{sec:acc_fmts}

Our main motivation for \glspl{issr} is accelerating sparse-dense linear algebra. As already outlined, indirection enables the pairwise streaming of \emph{nonzero-product operands} for any sparse tensor format whose major axis is given by two arrays: a \emph{value array} storing nonzeros, and an \emph{index array} storing their positions on the axis. We call such an array pair a \emph{sparse fiber}.

In addition to inherently representing sparse vectors, sparse fibers form the base of the \gls{csr}~\cite{eisenstat1977csr} and \gls{csc}~\cite{duff1989csc} matrix formats, concatenating the sparse rows or columns of a matrix as fibers and adding an array of pointers delimiting them. They also underlie the \gls{csf} tensor format~\cite{smith2015csf}, generalizing this idea.

\glspl{issr} therefore accelerate sparse-dense linear algebra with vectors, matrices, and general tensors in fiber-based formats; many format variations such as blocking and slicing can be supported through high-level iterators on the Snitch core.

\subsection{Core Product Kernels}
\label{sec:core_kernels}

To evaluate our streamer, we created a set of single-core sparse-dense product kernels operating on sparse fiber vectors and \gls{csr} matrices. We implemented dot product~(\gls{spvv}), \gls{csrmv}, and \gls{csrmm} kernels, each for 16- and 32-bit indices and in three variants:

\begin{itemize}
    \item \textsc{base}: Stock RISC-V optimized baseline
    \item \textsc{ssr}: RISC-V with \gls{frep} using \gls{ssr}
    \item \textsc{issr}: RISC-V with \gls{frep} using \gls{ssr} and \gls{issr}
\end{itemize}

Our \textsc{base} kernels are derived from the minimal indirection loop discussed in \cref{sec:intro}; we compile C kernels with \verb|gcc -O3| and hand-optimize the resulting assembly by instruction reordering and unrolling to minimize stalls. The \textsc{ssr} kernels stream the sparse vector values and are unrolled further where necessary to avoid stalls. Finally, we want to focus on our \textsc{issr} kernels which we will discuss in detail:

\paragraph*{SpVV}

\cref{lst:bg_ssr_dotp} outlines our \textsc{issr} \gls{spvv} kernel, with 16- and 32-bit index versions varying in the \gls{issr} configuration registers used. We aim to issue a continuous stream of \emph{fused multiply-add} instructions (\verb|fmadd.d|) with minimal overhead to maximize FPU utilization. We achieve this in three steps:

\begin{enumerate}[label=\roman*)]
    \item \emph{Setup}: we configure the \gls{ssr} to stream the sparse vector's values and the \gls{issr} to indirect into the dense vector's values at the sparse vector's indices. We enable redirection to our streamer and initialize a contiguous set of accumulator registers, here starting at \verb|ft2|, with zero.
    \item \emph{Compute}: We loop over an \verb|fmadd.d| instruction using \gls{frep}, streaming both operands from \glspl{ssr}. To prevent stalls due to read-after-write register dependencies, we use \gls{frep}'s \emph{register staggering} feature to auto-increment the accumulator register number on each iteration, maintaining several partial sums. Due to its lower peak utilization, the 32-bit kernel requires fewer accumulators. The \gls{frep} register staggering is described in more detail in~\cite{zaruba2020snitch}.
    \item \emph{Teardown}: we additively reduce our accumulators and store the result before disabling \gls{ssr} register redirection.
\end{enumerate}

\begin{listing}[!t]
\includegraphics{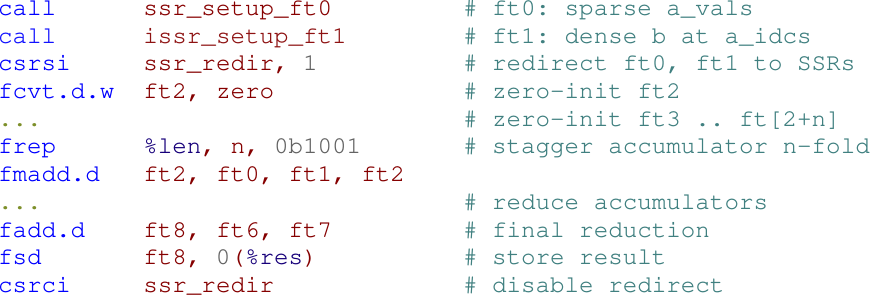}
\caption{\gls{issr}-accelerated \gls{spvv} kernel: \gls{ssr} \texttt{ft0} streams sparse vector \emph{a} and ISSR \texttt{ft1} values from dense vector \emph{b}.\vspace{-1em}}
\label{lst:bg_ssr_dotp}

\end{listing}

As the teardown contains only FPU subsystem instructions, the core is free to move on with execution after issuing all FPU instructions until data from the FPU is needed, enabling \emph{pseudo-dual-issue} operation; synchronization with the FPU can be enforced with a dummy register move if needed.

\paragraph*{CsrMV}

While we could reuse our \gls{spvv} kernel for each matrix row, we further optimize the \textsc{issr} \gls{csrmv} kernel to maximize FPU utilization:
\begin{itemize}
\item We stream the entire matrix fiber in single \gls{ssr} and \gls{issr} jobs, significantly reducing setup overhead.
\item We unroll the first few \verb|fmadd| in each row with branches to shorter reductions for rows with few elements, issuing an FREP loop and a full reduction only when necessary.
\end{itemize}

Our kernels use 32-bit row pointers, enabling broad scaling in rows. Our kernels primarily target \gls{csr} matrix multiplication from the left, but support power-of-two vector and arbitrary result strides, enabling multiplication of any power-of-two-strided dense axis with a \gls{csr} or \gls{csc} matrix from either side.

\paragraph*{CsrMM}

We multiply a \gls{csr} matrix with a power-of-two-column, dense row-major matrix to produce a dense row-major output. We reuse our \gls{csrmv} kernels, iterating on the dense matrix and result along their columns. Since we now iterate over dense data in a third-order loop, the overhead over our \gls{csrmv} kernels is small to negligible.

We support custom dense matrix and result strides, enabling products between row- and column-major matrices with \gls{csr} or \gls{csc} matrices from either side. The main restriction is that our index shifter requires a power-of-two stride on the indirected dense axis; in practice, this can be accommodated by tiling matrices into the \gls{tcdm} using the cluster \gls{dma}'s 2D transfers.

\subsection{Further Indirection Applications}
\label{sec:furth_appl}

While \glspl{issr} accelerate sparse-dense products and related operations like accumulating sparse onto dense tensors, indirection is a \emph{general-purpose} operation accelerating many more problems, some of which we want to highlight here:

\paragraph*{Codebook decoding}

\glspl{issr} can stream codebook-compressed data, representing arrays with repeated values as a series of indices pointing to a compact value array. Codebooks can be used in the quantization of image components~\cite{marcellin2002jpeg}, deep learning weights and activations~\cite{dave2020hardware}, or to compress nonzeros in sparse matrices with few unique values among other uses.

A single \gls{issr} can stream a codebooked vector; a streamer with two \glspl{issr} could accelerate sparse-dense products with codebook-compressed sparse values with near-identical code and performance as our existing kernels; this could strongly reduce the memory footprint of compressible tensors.

\paragraph*{Improved convolutions}

Accelerators often map dense convolutions to matrix multiplications through data transforms~\cite{dave2020hardware}. \glspl{ssr} can accelerate convolutions with \emph{rectangular} stencils without such transforms; \glspl{issr} could extend this capability to \emph{arbitrarily-shaped} sparse stencils by streaming an offset index array providing the stencil's shape and incrementing the data base address on the core.

\paragraph*{Scatter-gather streaming}

\glspl{issr} are, in effect, streaming \emph{scatter-gather units} as found in vector processors. As such, they could be used to accelerate scatter-gather algorithms including parallel radix sort\cite{zagha1991radix}, sparse matrix transpose~\cite{stathis2004transp}, and the densification of sparse tensors by nonzero scattering.
\section{Results}
\label{sec:results}

We will first present the performance of our kernels on a single core and in a cluster \gls{csrmv} implementation. We will then discuss the area and timing impacts as well as the energy efficiency benefits for our \gls{issr} streamer.

For all experiments, we obtained dense test tensors by sampling normally-distributed values. Sparse vectors were generated with normally-distributed values and uniformly-distributed indices given a fixed nonzero count and dimension. The sparse matrices used are all real-world-problem matrices from the SuiteSparse Matrix Collection \cite{davis2011suitesparse}: they have 2\si{\kilo} to 3.2\si{\kilo} columns, 1.3\si{\kilo} to 680.3\si{\kilo} nonzeros, varying aspect ratios, and cover various problem domains.

\subsection{Performance: Single Core}
\label{sec:results_cc}

To measure the architectural benefits of our \gls{issr} streamer without the influence of other system components, we evaluate our product kernels in RTL simulation of a single \gls{cc} by coupling it to ideal single-cycle instruction and two-port data memories. The latter behave similarly to the instruction cache and \gls{tcdm} in a cluster, except for misses and bank conflicts whose effects are included in \cref{sec:results_cl}.

\captionsetup[figure]{aboveskip=0em}
\begin{figure}[t]
    \centering
    \begin{subfigure}[b]{0.495\linewidth}
        \includegraphics[width=\linewidth]{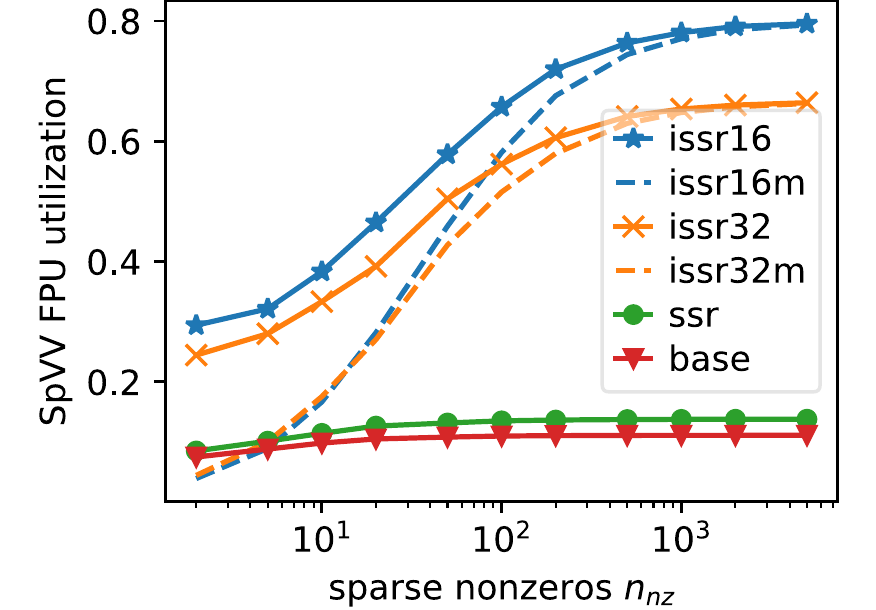}
        \caption{CC SpVV FPU utilizations}
        \label{fig:plot_cc_spvv}
    \end{subfigure}
    \begin{subfigure}[b]{0.485\linewidth}
        \includegraphics[width=\linewidth]{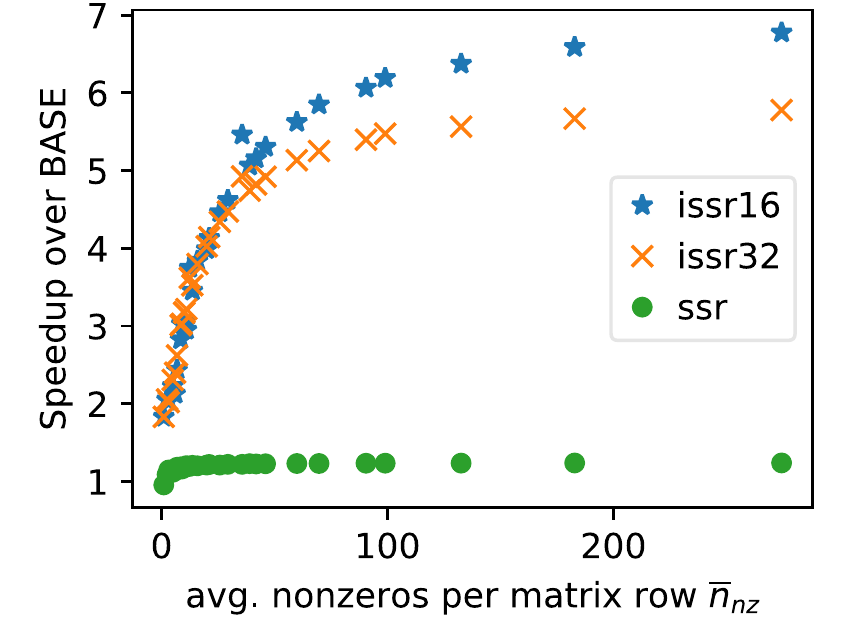}
        \caption{CC CsrMV speedups}
        \label{fig:plot_cc_spmv}
    \end{subfigure}
    \begin{subfigure}[b]{0.495\linewidth}
        \includegraphics[width=\linewidth]{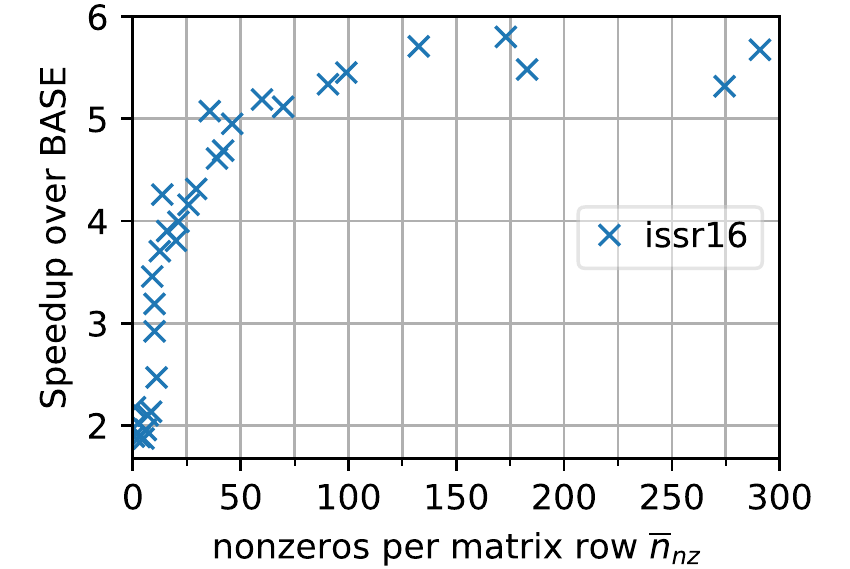}
        \caption{\textsc{issr} cluster CsrMV speedup}
        \label{fig:plot_cl_spmv}
    \end{subfigure}
    \begin{subfigure}[b]{0.485\linewidth}
        \centering
        \includegraphics[width=\linewidth,]{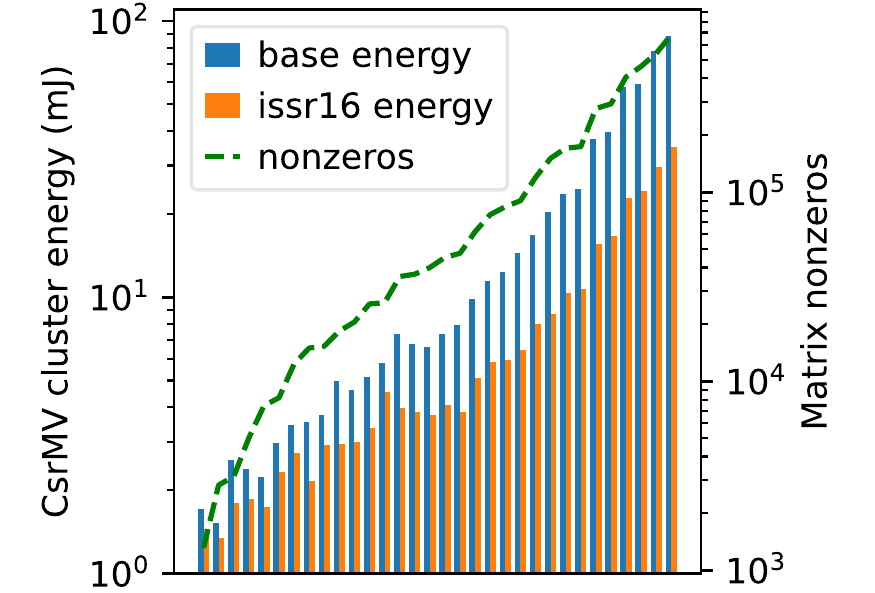}
        %\vspace{0.4cm}
        \caption{Cluster CsrMV energies}
        \label{fig:plot_cl_spmv_pow}
    \end{subfigure}
    \caption{Sparse-dense product kernel results}
    \label{fig:plots_perf_issr}
\end{figure}

\paragraph*{\gls{spvv}} \cref{fig:plot_cc_spvv} shows the FPU utilization for our dot product kernels, excluding load-store operations idling the datapath, against the sparse vector's nonzero count $n_{nz}$. We note that the runtime is independent of the dense vector's size as long as it fits into the \gls{tcdm}. Non-\textsc{issr} kernels perform identically for 16- and 32-bit indices as we cannot optimize halfword index loads. \textsc{issr} kernel \gls{fpu} utilizations are shown including and excluding accumulator reductions (\emph{m} suffix).

Both 16- and 32-bit \textsc{issr} kernels significantly outperform \textsc{base} and \textsc{ssr} kernels, increasing utilization by up to 4.7\x~and 5.6\x~over the latter and approaching their arbitration-imposed upper utilization limits of 0.8 and 0.67. Note that \textsc{base} and \textsc{ssr} kernels also approach their utilization limits of one multiply-accumulate every nine and seven cycles, respectively.

Because \textsc{issr}  kernels process each nonzero significantly faster, they require a proportionally higher $n_{nz}$ to overcome their \gls{ssr} setup and reduction overheads. For $n_{nz} < 5$, their reduction-free utilization is even lower than in non-\textsc{issr}  kernels, motivating our \gls{csrmv} row unrolling in such cases. As the 16-bit \textsc{issr} kernel needs more accumulators to sustain peak utilization, it outperforms the 32-bit variant only at higher $n_{nz}$.

\paragraph*{\gls{csrmv}}

\cref{fig:plot_cc_spmv} shows the speedup of \gls{csrmv} kernels over the \textsc{base} kernel against the average nonzeros per matrix row $\overline{n}_{nz}$, reflecting the inner loop iterations. Our experiments assume that the \gls{tcdm} is large enough to store the full matrix.

As for SpVV, our \textsc{issr} kernel speedups over \textsc{base} approach their theoretical 7.2\x~and 6.0\x~limits, requiring a higher $\overline{n}_{nz}$ than slower kernels to compensate for their overhead. Again, the 16-bit \gls{issr} kernel outperforms its 32-bit counterpart only past $\overline{n}_{nz} \approx 20$ due to its longer reduction.

\paragraph*{\gls{csrmm}} speedups and utilizations are near identical to the \gls{csrmv} kernels we use, even in edge cases: for a tiny sparse matrix like \emph{Ragusa18} with only 64 nonzeros, \gls{fpu} utilization changes by only \SI{0.12}{\percent} for a 2-column dense matrix.

\subsection{Performance: Cluster}
\label{sec:results_cl}

To evaluate performance on a system level, we implement multicore \gls{csrmv} on a Snitch cluster, reusing our single-core kernels, \emph{distributing rows} among cores, and employing a \emph{double-buffered} data movement scheme for the matrices using the cluster \gls{dma}. Our cluster is served by a 512-bit duplex main memory modeled as ideal. All data initially resides in main memory and results are written back to it.

\cref{fig:plot_cl_spmv} shows the speedup of the 16-bit \textsc{issr} kernel over the \textsc{base} kernel in cluster \gls{csrmv}. Speedups are significant even for $\overline{n}_{nz}=1$ at 1.9\x~and reach up to 5.8\x, sustaining over 5\x~for $\overline{n}_{nz}>50$. They follow the same trend as single-\gls{cc} \gls{csrmv} with \emph{reduced speedup} and \emph{stronger variations} from trends, which we attribute to multiple aspects:

Most notably, \gls{tcdm} \emph{bank conflicts}, accented by the random bank access patterns of indirection, lower peak \gls{fpu} utilization from 0.8 to 0.71. Additionally, the \emph{initial vector transfer} cannot be fully overlapped with computation, modulating speedups with the vector length. Finally, our \emph{row distribution} and \emph{double-buffering} schemes cannot fully prevent computation imbalance and overhead, and we encounter some instruction cache stalls.
Despite these inherent parallelization challenges, \glspl{issr} enable extreme efficiency gains with minimal area overhead: \emph{eight} cores with \glspl{issr} achieve the same peak floating-point throughput as \emph{46} cores running \textsc{base}.

\subsection{Area and Timing}
\label{sec:results_synth}

We synthesize the \gls{ssr} and \gls{issr} streamers using Synopsys \emph{Design Compiler} for GlobalFoundries’ \SI{22}{\nano\meter} FD-SOI technology, targeting the SSG corner at \SI{-40}{\celsius} with low-$V_t$ cells, \SI{0.72}{\volt} supply voltage, and no back-biasing. A \SI{1}{\giga\hertz} clock and \SI{100}{\pico\second} IO delays were constrained. Our hardware configuration has one \gls{ssr} and one \gls{issr} with five data FIFO stages, 18-bit indices and addresses, and four affine loops.

Compared to the \gls{ssr}, the \gls{issr}'s longest path increased from \SI{301}{\pico\second} to \SI{425}{\pico\second}, still easily meeting Snitch's \SI{1}{\giga\hertz} clock target.
The \gls{issr} streamer's hierarchical area distribution is denoted in \cref{fig:arch_hssr_streamer}. Our \gls{issr} is \SI{4.4}{\kGE} or \SI{43}{\percent} larger than the equivalently parameterized \gls{ssr}, incurring only a negligible \SI{0.8}{\percent} area overhead in our eight-core Snitch cluster compared to providing only \glspl{ssr}.

\subsection{Energy and Power}
\label{sec:results_power}

We estimate the power consumption of a Snitch cluster running \gls{csrmv} with \textsc{base} and  16-bit \textsc{issr} kernels on our matrix set, targeting the TT corner in  GlobalFoundries’ \SI{22}{\nano\meter} FD-SOI at \SI{1}{\giga\hertz}.
We use Synopsys \emph{Design Compiler} to topographically synthesize our cluster and Synopsys \emph{PrimeTime} to estimate power for the low- and high-efficiency matrices \verb|G11| and \verb|G7|, then scale dynamic power with hardware component utilizations measured in RTL simulation for all matrices.

\cref{fig:plot_cl_spmv_pow} shows the total energy for \gls{csrmv} using both kernels for each matrix. While the peak average cluster power is predictably lower for the \textsc{base} kernel (\SI{89}{\milli\watt} vs. \SI{194}{\milli\watt}), \glspl{issr} achieve energy efficiency improvements of up to 2.7\x~(\SI{142}{\pico\joule} to \SI{53}{\pico\joule} per \verb|fmadd|).

\section{Related Work}
\label{sec:relwork}

\paragraph*{General-purpose processors} \glspl{issr} are most related to \emph{scatter-gather} hardware in vector processors, which is slowly adopted by superscalar out-of-order architectures:
Intel's \gls{knl} \cite{sodani2016knl} and Arm's Scalable Vector extensions \cite{stephens2017sve} introduce \gls{simd} scatter-gather. However, \gls{simd} vectors are only \emph{a few elements} wide, severely limiting their applicability to sparsity compared to \glspl{issr}.
Xie et al. \cite{xie2018cvr} optimize \gls{spmv} on \gls{knl} with their own sparse format, improving \gls{simd} lane and cache usage efficiency over the state of the art. Still, they use at most \SI{0.7}{\percent} of their system's peak double-precision compute, 70\x~less than what we achieve in a Snitch cluster with \glspl{issr}.

\paragraph*{GPUs} sparse-dense operations on GPUs are often approached in \emph{software} through efficient algorithms and sparse formats:
Shi et al. \cite{shi2020sdmvgpu} accelerate sparse matrix-matrix products with their own matrix format and algorithm.
Merrill et al. \cite{merrill2016mergegpu} propose \gls{csrmv} using a merge-based decomposition to improve performance consistency.
Nvidia's \emph{cuSPARSE} library \cite{nvidia2018cuda} provides optimized sparse problem kernels. We evaluate their \gls{csrmv} kernels from CUDA Toolkit 10.0 on a GTX 1080 Ti GPU (Pascal GP104 architecture, FP32 and FP64) and a Jetson AGX Xavier (Volta architecture, FP32 only). We reuse the matrices from our evaluation to profile 100 consecutive kernel runs using \emph{nvprof} \cite{nvidia2018cuda}. For FP32, both platforms exhibit high peak \gls{sm} occupancy (\SI{87}{\percent} and \SI{96}{\percent}), but low peak floating-point utilizations (\SI{0.75}{\percent} and \SI{2.1}{\percent}) on active \glspl{sm}, suggesting low thread parallelism among warps. FP64 shows similar \gls{sm} occupancies with a notably higher peak floating-point utilization (\SI{17}{\percent}), likely due to the 32\x~fewer FP64 cores per \gls{sm}. Still, peak utilization is 2.8\x~lower than in a Snitch cluster with \glspl{issr}.

More recently, GPU \emph{hardware} targeting sparsity has been proposed: Zhu et al. \cite{zhu2019sparsetensorcore} present an algorithm and tensor core modifications for efficient sparse neural network inference. Nvidia's A100 architecture \cite{nvidia2020a100} supports \emph{structured} sparsity, efficiently handling up to two zeros in every four values. This imposed structure covers only a small subset of our \glspl{issr}' capabilities, which efficiently handle a much wider range of sparsities ($\gg\;$\SI{50}{\percent}) and random accesses into a full \gls{tcdm}.

\paragraph*{Hardware accelerators} Sparse accelerators often target specific domains like \gls{dl} \cite{dave2020hardware}.
Direct comparisons to our work are hard since most accelerators use highly application-specific data precisions and formats and are incapable of general-purpose computation. Nevertheless, a programmable Snitch cluster with \glspl{issr} can implement many highly specialized computation schemes used in hardwired accelerators.
Han et al. \cite{han2016eie} keep entire pruned, weight-shared \gls{dl} models in SRAM, minimizing DRAM access energy; \glspl{issr} work with any \gls{tcdm} stationarity scheme and enable codebooked weight reuse.
Hedge et. al \cite{hedge2019extensor} hierarchically intersect sparse tensor indices to avoid zero products and their memory traffic; a Snitch core could intersect upper tensor axes while its \gls{issr}-fed \gls{fpu} processes nonzeros.

\section{Conclusion}
\label{sec:conc}

In this work, we extend \glspl{ssr} for \emph{streaming indirection}, creating a backward-compatible \gls{issr} which enables efficient sparse-dense tensor products with sparse-fiber-based formats including \gls{csr} and \gls{csf}. We demonstrate it in the RISC-V Snitch system by presenting efficient dot product, \gls{csrmv}, and \gls{csrmm} kernels, and propose further indirection applications.

We evaluate our \gls{issr} streamer by comparing \gls{issr} product kernels to optimized-baseline and \gls{ssr}-only kernels on a single \gls{cc}; we improve peak dot product FPU utilization from 0.11 (\textsc{base}) or 0.14 (\textsc{ssr}) to 0.67 and 0.80 for 16- and 32-bit indices, and enable \gls{csrmv} speedups of up to 7.2\x~over an optimized baseline. In a multicore cluster \gls{csrmv} implementation, speedups over our baseline reach up to 5.8\x~despite work sharing, vector transfer, and bank conflict overheads. The default \gls{issr} is only \SI{4.4}{\kGE} or \SI{43}{\percent} larger than an \gls{ssr} with a longest path of \SI{425}{\pico\second} in GF22FDX technology, but improves cluster \gls{csrmv} energy efficiency by up to 2.7\x~over our baseline. A Snitch cluster with \glspl{issr} provides wider sparse-dense support and higher floating-point utilizations (70\x~and 2.8\x) than recent CPUs and GPUs, yet is flexible enough to adopt key innovations of sparse accelerators in software.

\bibliographystyle{IEEEtran}
\bibliography{IEEEabrv,content/bibliography}

\end{document}